\documentclass[prb,showpacs,preprint,amsmath]{revtex4}
\usepackage{graphicx}% Include figure files
\usepackage{dcolumn}% Align table columns on decimal point
\usepackage{bm}% bold math

\newcommand{\mathcommand}[3][0]{\newcommand{#2}[#1]{\ensuremath{#3}}}
\newcommand{\be}{\begin{equation}}
\newcommand{\ee}{\end{equation}}
\mathcommand[1]{\ket}{\left| #1 \right\rangle}  %a Ket
\mathcommand[1]{\bra}{\left\langle #1 \right|}  %a Bra
\mathcommand[2]{\ketbra}{
    \left| #1 \right\rangle\!\!\left\langle #2 \right|} %an outer product

\mathcommand[2]{\ts}{#1_{\text{#2}}}
\mathcommand{\te}{\text{e}}
\mathcommand{\thole}{\text{h}}
\mathcommand{\nodag}{{\phantom{\dag}}}

\newcommand{\NII}{{
    National Institute of Informatics,
    2-1-2 Hitotsubashi, Chiyoda-ku,
    Tokyo 101-8430, Japan
}}

\newcommand{\Ginzton}{{
    $^{2}$Edward L. Ginzton Laboratory,
    Stanford University,
    Stanford, California 94305-4088, USA
}}

\newcommand{\Paderborn}{{
    $^{1}$Dept. of Physics,
    University of Paderborn,
    Warburger Str. 100, 33098
    Paderborn, Germany
}}

\begin{document}
\title{Lasing of donor-bound excitons in ZnSe microdisks}
\author{A. Pawlis$^{1,2}$}
\email[Electronic address: ]{pawlis@stanford.edu}
\author{M. Panfilova}
\author{D. J. As}
\author{K. Lischka}
\affiliation\Paderborn
\author{K. Sanaka}
\author{T. D. Ladd}
\author{Y. Yamamoto}
\affiliation\Ginzton\affiliation\NII

\begin{abstract}
Excitons bound to flourine atoms in ZnSe have the potential for
several quantum optical applications. Examples include optically
accessible quantum memories for quantum information processing and
lasing without inversion.  These applications require the
bound-exciton transitions to be coupled to cavities with high
cooperativity factors, which results in the experimental observation
of low-threshold lasing. We report such lasing from fluorine-doped
ZnSe quantum wells in 3 and 6~micron microdisk cavities.
Photoluminescence and selective photoluminescence spectroscopy
confirm that the lasing is due to bound-exciton transitions.
\end{abstract}

  % Quantum Computation

\maketitle

Quantum interference of the two optical pathways of an optical
$\Lambda$-system, in which two long-lived ground states are
optically coupled to a single excited state, provides a powerful
mechanism for a number of useful applications.  These applications
include lasing without inversion \cite{lwi}, electromagnetically
induced transparency \cite{hfi90}, and optically addressable quantum
memory for quantum information processing
\cite{czkm97,yls05,ctsl05,wv06,susan,hybridnjp}.  The development of
scalable technologies based on these effects would be facilitated by
solid-state implementations, which are generally more difficult than
atomic demonstrations due to optical inhomogeneity and decoherence.

In semiconductors, donor impurities and charged quantum dots both
provide promising realizations of such $\Lambda$-systems.  The
long-lived ground states are provided by the bound electron spin in
high magnetic field, while the optically excited state is formed by
the lowest-energy donor-bound exciton or trion state.  For
applications involving quantum memory, however, quantum dots have
the disadvantage of severe inhomogeneity in optical transition
frequencies and electron magnetic moments. Such inhomogeneity may
inhibit the ability to incorporate many optically interacting
quantum-dot-based $\Lambda$-systems into a scalable
quantum-information-processing device.  In contrast, the potentials
provided by donor impurities in a perfect crystal are very
homogeneous, as demonstrated by high-quality samples of
GaAs~\cite{Karasyuk94a} and Si~\cite{thewalt}. Homogeneous ensembles
of donor-bound excitons in bulk GaAs have been demonstrated to
exhibit coherent-population trapping~\cite{Kai-Mei}, which is one
effect exhibiting quantum coherence between the two emission
pathways of the optical $\Lambda$-system.

Donor-bound excitons in ZnSe are of particular interest because ZnSe
may be isotopically purified to feature only spin-0 substrate
nuclei, avoiding the disadvantage of decoherence limited by the
dynamics of nuclear spins in \textsc{iii-v} semiconductor
systems~\cite{desousa}. Silicon-based systems also have the
potential to overcome nuclear decoherence~\cite{lyon}, but silicon
is optically dark due to its indirect bandgap.   In a \textsc{ii-vi}
system such as ZnSe, the nuclear spins may be entirely removed from
the substrate as in the case in silicon, and the donor-bound exciton
emission is optically bright as in the case of GaAs.   The fluorine
donor is of particular interest because it provides a 100\% abundant
spin-1/2 nucleus, which may be employed for long-lived storage of
quantum information.

ZnSe is a wide-bandgap semiconductor which emits light in the blue
region ($\lambda \approx 440$~nm), where lasing is traditionally
difficult to obtain.  Early studies of lasing in ZnSe for the
purpose of blue lasers were limited because the material is too
soft to be used in high-current or high-excitation regimes, so
that heating effects degenerate the active medium~\cite{chao1997,okuyama2000}.
However, low-threshold lasing is achievable when a high-$Q$,
low-volume optical microcavity introduces a high ratio of stimulated
emission to spontaneous emission, known as the
$\beta$-factor~\cite{ymb91}.  Such low-threshold lasing is seen in
recent developments in microcavity quantum dot lasers~\cite{strauf}.
Further, if the lasing medium is provided by discrete donor-bound
exciton transitions rather than the continuum of free-excitons, the
coherence provided by the spin-ground states of the donor-spins
could allow the possibility of lasing without inversion~\cite{lwi},
further decreasing the input power requirements.

Lasing in ZnSe donor-bound excitons may be particularly useful as a
component in quantum information processing devices.
%If the $^{19}$F
%nucleus is used for quantum memory in such devices, its spin-state
%could in principle be resolved optically if the donor-bound emission
%were comparable or narrower than the hyperfine
%splitting~\cite{nucmeas}.
%At sufficiently high lasing powers, such a
%narrow line may be provided at the Schawlow-Townes limit. Further,
The design of quantum computers~\cite{susan} or quantum
repeaters~\cite{hybridnjp} based on donor-bound excitons in optical
microcavities requires a low-noise source laser nearly resonant the
bound-exciton transitions used for qubit initialization, control,
and read-out.  A laser based on the same medium as the qubit
provides promising pathways for device integration.

Bound-exciton emission in ZnSe was observed many years
ago~\cite{merz}, and the optical transitions of single, isolated
ZnSe acceptors were recently measured~\cite{Strauf2002}.  However,
coherent applications involving $\Lambda$-systems are more likely to
be achieved from donors, due to the longer coherence time of
electrons over holes. The $^{19}$F-donor has been studied earlier in
bulk samples~\cite{Pawlis2006}.  In this Letter, we report evidence
of $^{19}$F-donors optically coupled to microcavities exhibiting
low-threshold lasing of the donor-bound exciton transitions.

\begin{figure}
\centering
\includegraphics[width=3in]{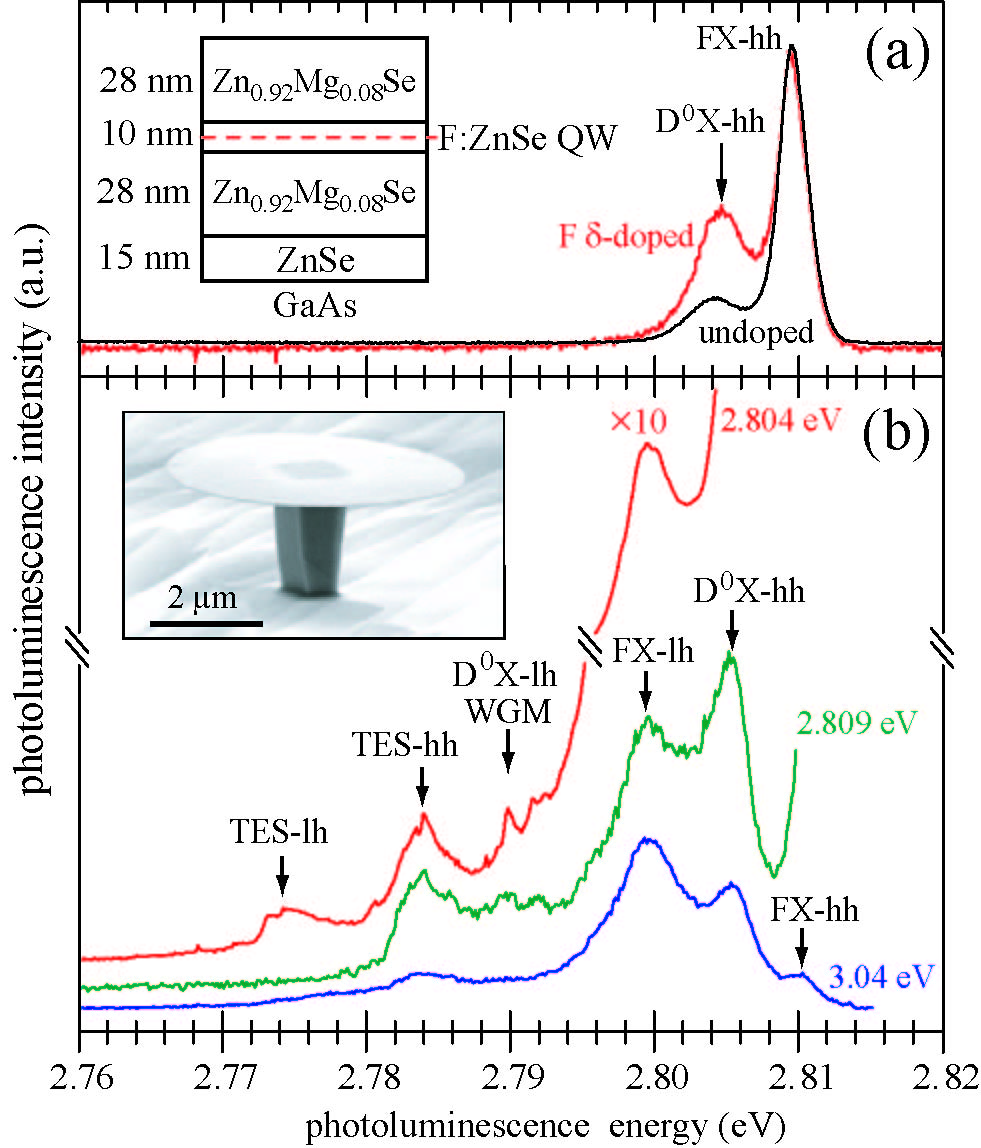}
\caption{\label{PLBoth} PL and SPL spectra taken at 5~K.
    (a)
PL for unstructured F:ZnSe $\delta$-doped QW (red curve) and undoped
QW (black curve).
Inset:~The material composition of the unstructured sample.
    (b)
Above-band PL (blue curve) and SPL spectra of a 6~$\mu$m-disk taken
under resonant excitation of the D$^0$X-hh peak (red curve, laser
energy 2.804 eV) and the FX-hh peak (green curve, laser energy 2.809
eV). Inset:~Scanning electron micrograph of a 6~$\mu$m disk.
    }
\end{figure}

Previous work on microcavity lasers in the wide-bandgap
\textsc{ii-vi} semiconductors have focused on lasers based on CdSe
quantum dots~\cite{worschech2006, renner06} or ZnCdSe quantum wells
(QW's)~\cite{hovinen93} confined by quaternary alloys of ZnMgSSe and
ZnCdSSe, which are lattice-matched to a GaAs substrate.  To reduce
alloy fluctuations, we confined a ZnSe QW in ternary ZnMgSe barriers
with low Mg-concentration.  Lasing due to confined free-excitons
seeded by excitonic molecules has been previously observed in
similar systems~\cite{kreller95,kozlov96}. Here we show lasing
directly on discrete, bound-exciton transitions in samples with a
fluorine $\delta$-doped ($^{19}$F:ZnSe) layer at the center of the
QW.

The ZnMgSe/ZnSe/$^{19}$F:ZnSe/ZnSe/ZnMgSe samples were grown by
molecular beam epitaxy on GaAs-(001) substrates.  For optimal
interface properties, a 15~nm buffer of undoped ZnSe was first
deposited on the substrate, followed by 28~nm of ZnMgSe with Mg
content of about 8\%. The ZnSe QW was 10~nm thick.  For doped
samples, the QW was $\delta$-doped in the center using a ZnF$_2$
evaporation cell with a net molecular flux of approximately $2\times
10^9$~cm$^{-2}$s$^{-1}$, equivalent to a sheet donor concentration
of $8\times 10^{9}$~cm$^{-2}$.

Photoluminescence (PL) spectra of unstructured $^{19}$F-doped and
undoped samples were measured at 5~K, with excitation provided by a
blue light-emitting diode of wavelength 365~nm
[Fig.~\ref{PLBoth}(a)]. The PL spectra of these strained samples
show two peaks.
%The emission at 2.810~eV corresponds to
%recombination of heavy-hole free excitons (FX-hh) from the
%compressively strained ZnSe; this energy is consistent with the
%finite barrier model reported in Refs.~\onlinecite{teo94} and
%\onlinecite{chung97}.
The emission at 2.810 eV corresponds to recombination of heavy-hole
free excitions (FX-hh) from the compressively strained ZnSe QW
(lattice mismatch of f(ZnSe)=$-0.25\%$); this energy is consistent
with the finite barrier model in Refs. 27 and 28.

Light-hole free-exciton (FX-lh) emission at 2.829~eV is not observed
at low temperature. The lower energy peak seen at 2.804~eV in
Fig.~\ref{PLBoth} is due to the heavy-hole donor-bound exciton
(D$^0$X-hh) recombination, as confirmed by the expected 6~meV
separation and the 3-fold increase in this line provided by the
fluorine $\delta$-doped (red curve) versus the undoped sample (black
curve).

The widths of these peaks indicate inhomogeneous broadening, likely
due to strain effects. However, the broadening is still less than
typically seen in ensembles of quantum dots, and may be useful for
the isolation of individual impurities.

Micro-disks with diameters of 3~$\mu$m and 6~$\mu$m were defined by
photolithography, reactive ion etching, and wet etching of the
underlying GaAs substrate.  The inset in Fig.~\ref{PLBoth}(b) shows
a scanning electron micrograph picture of such a microdisk.  The
absence of cracks and lateral deformation indicates a homogeneous
release of strain along the radial direction of the disks; however,
considering the volume ratio of ZnMgSe vs. ZnSe in the disks, the
ZnSe QW is likely to be tensile strained on ZnMgSe in the periphery
of the disks.  This results in a substantial band gap narrowing
effect along the radial direction of the disk from the center to the
edge which enhances the charge carrier transfer to the periphery,
where the optical whispering gallery modes (WGMs) are localized. As
a result of the tensile strain, the relative positions of the FX-lh
and FX-hh energies are reversed and both states contribute to the
PL.

Bandgap narrowing and strain relaxation are confirmed by the
corresponding PL-spectrum in Fig.~\ref{PLBoth}(b) taken from a
microdisk with 6~$\mu$m diameter. The modified transition energies
are found by fitting the above-band PL to 5 Gaussian peaks.
%These
%energies are again consistent with the finite barrier model reported
%in Refs.~\onlinecite{teo94} and \onlinecite{chung97} with a
%remaining average strain of the ZnMgSe barriers of 18\%.
These energies are again consistent with the finite barrier model
reported in Refs.~\onlinecite{teo94} and \onlinecite{chung97} with a
remaining induced tensile lattice mismatch of the quantum well of
f(ZnSe)=$+0.2\%$ in the etched structure.

The FX-hh and D$^0$X-hh transitions at 2.81~eV and 2.804~eV are only
slightly shifted and broadened; these peaks come from PL around the
center of the disk, where the material remains fully strained.  No
cavity modes are ever observed in this region of the spectrum.  The
D$^0$X-hh luminescence is now stronger than the FX-hh luminescence
due to surface recombination of the FX, and satellite transitions
are now visible at a position of 21~meV beneath the D$^0$X-hh
luminescence.  This energy is in agreement with the calculated
position of the two electron satellite (TES) of the
D$^0$X~\cite{dean}. In the relaxed periphery, the FX-lh emission
moves to the lower energy position of about 2.799~eV; the D$^0$X
transitions again appear 6~meV lower than their corresponding FX
transitions, but are substantially broadened due to the strain
gradient across the disk.

The association of the spectral peaks in structured samples is also
confirmed by selective PL (SPL), in which the FX-hh
transition and the D$^0$X-hh transition are resonantly excited by a
scanning narrow-band Ti:sapphire ring laser doubled by an LBO
crystal in a high-finesse cavity. Fig.~\ref{PLBoth}(b) shows the SPL
spectra of a 6~$\mu$m micro-disk. By resonant excitation of the
FX-hh transition at 2.809~eV (green curve),  the TES-hh lines are
enhanced relative to above-band excitation. Resonant excitation of
the D$^0$X-hh peak (red curve) results in the appearance of a new
satellite peak at 2.774~eV which we identify as TES-lh transitions.
These observations of TES lines in SPL strongly confirm the expected
contribution of F-donor-bound excitons to the band-edge luminescence,
as opposed to emission from quantum dots due to surface fluctuations
of the QW.  We also repeatedly observe a sharp peak at
2.79~eV when resonantly exciting the D$^0$X-hh line.  This peak
corresponds to a WGM of the microdisk.

\begin{figure}
\centering
\includegraphics[width=3in]{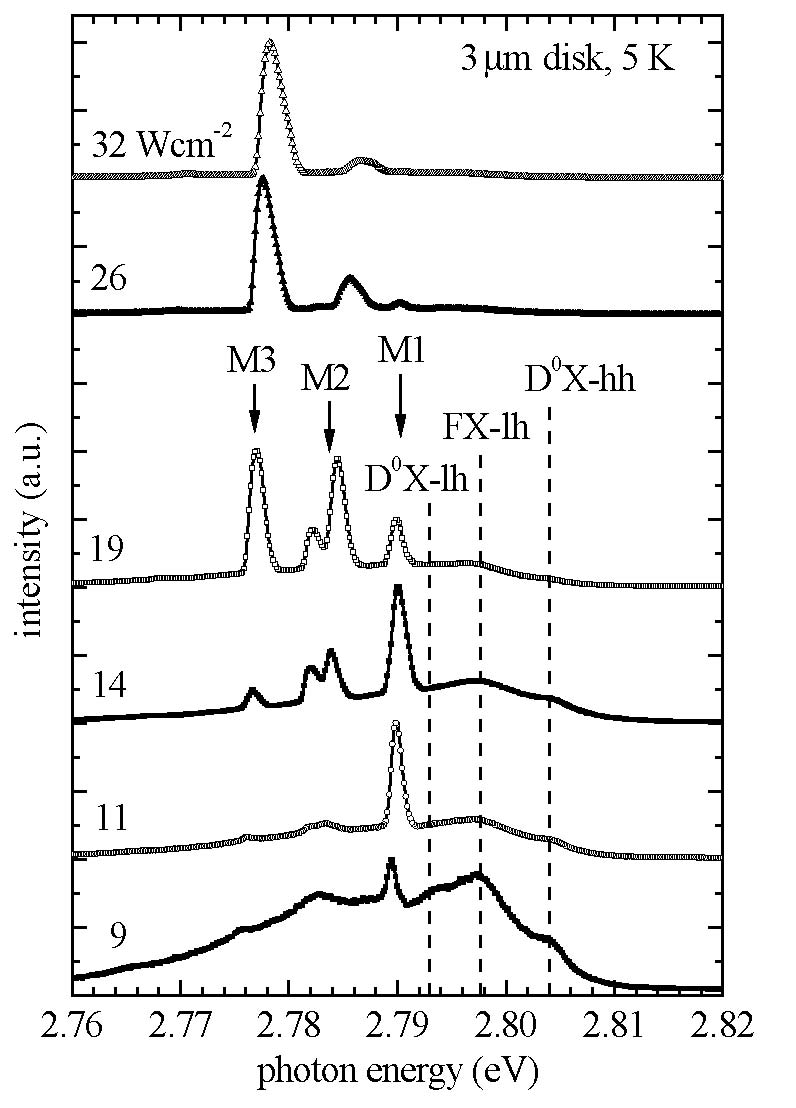}
\caption{\label{lasing} Power-dependent PL spectra and lasing of a
microdisk with 3~$\mu$m diameter. The excitation density was varied
between 9 and 32~W~cm$^{-2}$. For low excitation the WGM (mode M1)
emerges at 2.790~eV. The energies of the relevant hh- and lh-
transitions are indicated by dashed lines. Further modes (M2 and M3)
are observed at moderate excitation densities.}
\end{figure}

\begin{figure}
\centering
\includegraphics[width=3in]{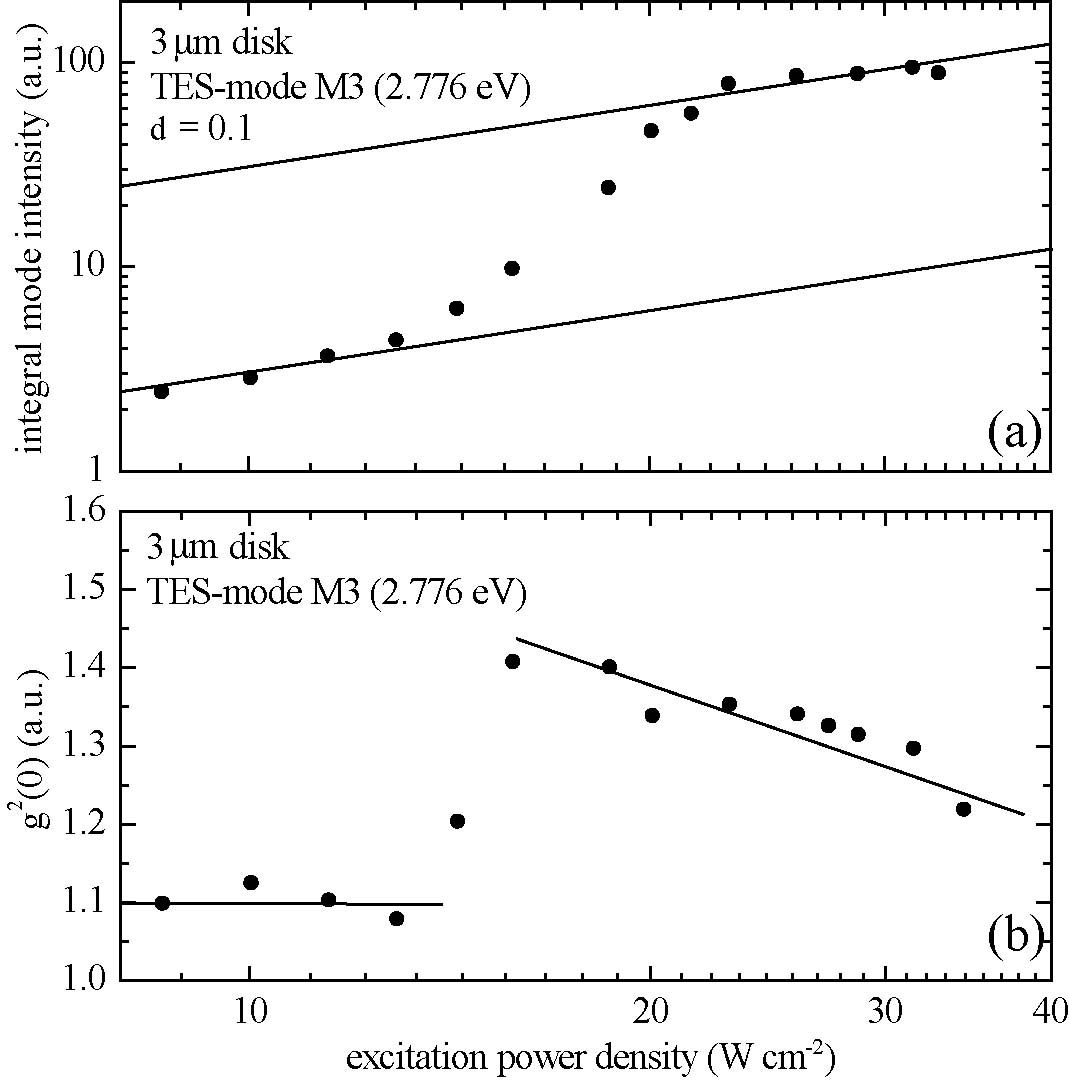}
\caption{\label{HBT} (a) Integrated intensity of the TES lasing mode
M3 at 2.776~eV versus the excitation power density. (b)
Power-dependent second-order photon correlation measured on mode
M3.}
\end{figure}

The WGM modes are enhanced with high-power pumping.
Figure~\ref{lasing} shows a series of PL spectra taken from a
micro-disk with 3~$\mu$m diameter and measured with different
average excitation intensities varying between 9 and 32~W~cm$^{-2}$.
For small power densities exceeding 5~W~cm$^{-2}$, superlinear
increase of a sharp peak at 2.790~eV (mode M1 in Fig.~\ref{lasing})
is observed. This peak corresponds to a WGM of the micro-disk
excited by D$^0$X-lh emission. The linewidth indicates a cavity-$Q$
of about 2500.   Additional modes emerge at energies of 2.783~eV and
2.785~eV (mode M2) and at 2.776~eV (mode M3) for pump power
densities larger than 12~W~cm$^{-2}$.  These modes are too closely
spaced to correspond to the free spectral range of the lowest order
WGMs; the multiple modes seen are likely due to higher order radial
modes of the disk.  The strain gradient and slight warping of these
disks prevent a simple quantitative evaluation of their spacing.
Mode M2 consists of two competing modes close to each other which
might stem from anisotropy of the disk shape.  At higher excitation
density the overall intensity of mode M1 and mode M2 is transferred
to mode M3, which is resonant with the TES transition energies of
the bound exciton states. The WGMs are only observed in the lower
energy part of the spectra corresponding to emission from donor
bound excitons and their related transitions, resulting from the
bandgap narrowing at the periphery of the disk.

The peaks corresponding to WGMs are not visible in low-power PL
using either a continuous wave 405~nm diode laser or the doubled
408~nm output of a mode-locked Ti:sapphire laser. We believe this is
because absorption due to lower-energy transitions in the
inhomogeneously broadened distribution damps the cavity-$Q$.
However, at higher excitation powers, these lower-energy transitions
become saturated and the cavity modes appear.

Figure~\ref{HBT}(a) shows the power dependence of mode M3 as a
function of the pump power. A clear lasing threshold is seen at
about 15~W~cm$^{-2}$.  This threshold is substantially smaller than
observed in previous \textsc{ii-vi}-semiconductor microdisk lasers
based on quantum dots~\cite{renner06}.  The ratio of the slope of
output power vs. input power indicates the fraction of spontaneous
emission coupled into the lasing mode to be $\beta = 0.1$.  Lasing
was also observed in 6~$\mu$m micro-disks, revealing threshold power
densities larger than 100~W~cm$^{-2}$ for the TES transition with a
corresponding $\beta$-factor of about 0.03.

Lasing was confirmed by the power dependence of the second order
photon correlation function $g^{(2)}(0)$, measured at the TES lasing
mode M3 with a standard Hanbury-Brown Twiss (HBT) experiment and
shown in Fig.~\ref{HBT}(b). Below threshold the spontaneous emission
into mode M3 follows a nearly Poisson distribution with $g^{(2)}(0)
\approx 1.1$. Near threshold we observe significant photon bunching,
manifested as $g^{(2)}(0) > 1$. Such bunching occurs as the
coherence length of the photons increases near threshold but before
the fully coherent output of above-threshold lasing.  Far above
threshold (powers exceeding 30~W~cm$^{-2}$), the HBT signal
converges back to $g^{(2)}(0) \approx 1$, indicating the generation
of a coherent state.

The observed $\beta$ factors allow us to estimate the cooperativity
factors of the microdisk cavities. In microdisks such as these that
are very wide and very thin relative to the emission wavelength, the
spontaneous emission rate into all cavity modes including unguided
modes sum to very nearly the spontaneous emission rate in the bulk.
Therefore, the $\beta$ factor for one of the WGMs is approximately
$\beta = (1+C)\beta_0$, where $\beta_0$ is the ratio of the emission
rate into a vanishing-$Q$ WGM normalized by the bulk emission rate
and the cooperativity factor $C$ manifests as the Purcell effect.

We estimate the mode volume $\beta_0$ using the approximation
presented in Ref.~\onlinecite{cch94} in the large-radius limit.  We
find that for disks as thin as these, the WGM's are quite wide,
allowing about 500 impurities to contribute to the lasing in a
3~$\mu$m disk.  This model yields $\beta_0\approx 0.02$, for an
expected cooperativity factor of about $C=4$, which is consistent
with estimates based on a $Q$ of 2500 and the calculated mode
volume.  Both $C$ and $\beta_0$ scale inversely with the disk
radius, consistent with the smaller observed $\beta$ factor of the
6~$\mu$m disk.

The appearance of low-threshold lasing in $\delta$-doped
$^{19}$F:ZnSe QW micro disks indicates that small ensembles of
$^{19}$F-Donors have been successfully coupled to a cavity with
cooperativity factors high enough to allow some schemes for quantum
communication and quantum computation~\cite{hybridnjp}.  The
reduction of inhomogeneous broadening by preserving the uniform
strain acquired during MBE growth may allow such applications as
lasing without inversion.   Further work will involve the isolation
of individual impurities in smaller microcavities, the measurement
of the donor-spin coherence times, and the measurement of single
$^{19}$F nuclear spin states.

This work was supported by \textsc{nict}.  We thank Kristiaan De
Greve and Shinichi Koseki for valuable discussions and experimental
assistance.

%\bibliographystyle{apsrev}
%\tiny
\bibliography{ZnSeCQED}
%GATHER{ZnSeCQED.bib}

\end{document}